# Coherently excited nonlocal quantum features using polarization-frequency correlation between quantum erasers

Byoung S. Ham[1,2]
[1]School of Electrical Engineering and Computer Science, Gwangju Institute of Science and Technology,
123 Chumcangwagi-ro, Buk-gu, Gwangju 61005, South Korea
[2]Qu-Lidar, 123 Chumcangwagi-ro, Buk-gu, Gwangju 61005, South Korea
(Submitted on Aug. 11, 2023, bham@gist.ac.kr)

**Abstract:**
Photon indistinguishability is an essential concept to understanding "mysterious" quantum features from the viewpoint of the wave-particle duality in quantum mechanics. The physics of indistinguishability lies in the manipulation of quantum superposition between orthonormal bases of a single photon such as in a quantum eraser. Here, a pure coherence approach is applied for the nonlocal correlation based on the polarization-frequency correlation of Poisson-distributed coherent photon pairs to investigate the role of measurements. For this, a gated heterodyne-detection technique is adopted for coincidence measurements between space-like separated delayed-choice quantum erasers, resulting in an inseparable basis product between them. For this coherently induced inseparable basis product, polarization-frequency correlated photon pairs are selectively measured through a dc-cut ac-pass filter to eliminate unwanted group of polarization-product bases. Finally, the Bell inequality violation is numerically confirmed for the coherence solutions of the nonlocal correlation.

**Introduction**
In quantum mechanics, measurements can retrospectively affect the original quantum state, violating the cause-effect relation of classical physics [1,2]. Such measurement-based quantum mechanics has been intensively studied for various delayed-choice experiments over the last several decades [3-12]. Not only a single photon-based quantum superposition [4-7] but also two photon-based quantum entanglement [10-13] has been intensively studied for the mysterious quantum features. Potential quantum loopholes in such a quantum system have been closed for detection [13,16], locality [14], sampling [13,15,16], and free will [17] parameters. The nonlocal correlation between space-like separated photons results in the violation of local realism [18-24], where the local realism is for predetermined physical laws that cannot be influenced by the nonlocal measurements as claimed by Einstein and his colleagues [18]. Thus, nonlocal realism should defy classical reality via the so-called "spooky action at a distance" [20]. However, this nonlocal correlation violating local realism has not been clearly understood yet. To achieve the 'mysterious' nonlocal correlation, nonclassical light sources have been used [13-30].

Here, a contradictory idea is presented to firmly understand the nonlocal correlation using classical means of coherence optics. For this, a coherently excited delayed-choice quantum eraser is investigated for the inseparable basis-product relation of local parameters. As usual, a noninterfering Mach-Zehnder interferometer (NMZI) with orthogonal polarization bases of a photon is applied for the quantum eraser [4-7]. For the coherence analysis, a pair of symmetrically frequency-detuned coherent photons is randomly excited from an attenuated laser using synchronized acousto-optic modulators (AOMs), resulting in polarization-frequency correlation through a polarizing beam splitter (PBS) in the NMZI. The polarization-frequency correlated photon pairs between two independent NMZIs are applied for the selective measurement of orthogonal polarization-product bases only via a gated detection of the heterodyne signals. For the gated detection, a dc-cut as-pass filter is inserted for the selective measurement process to eliminate the same frequency-product bases. To freeze the time window of the heterodyne signals, the temporal resolution of a photon detector must be much better than the inverse of the modulation bandwidth of AOMs (see Section A of the Supplementary Materials).

Over the last few decades, the quantum feature of Franson-type nonlocal correlation [24] has been experimentally demonstrated for the violation of local realism [25-30], whose interferometer satisfies the NMZI for ensemble photons [31]. Recently, the Franson correlation has been coherently investigated for selective measurements of product bases [31]. In Ref. [31], the nonlocal property of inseparable product basis is understood as a result of selective measurements at the cost of 50 % event loss. Similarly, the role of the ac-pass filter in the present scheme of the gated detection is to selectively choose a certain group of polarization-product bases at a frozen time to avoid a time averaging effect (see Section A of the Supplementary Materials). Wheeler's delayed-choice experiments [3-12] have also been investigated for the post-measurement control of polarization bases in a pure coherence approach, resulting in the same causality violation [32], even in a macroscopic regime [33]. As in the Franson-correlation scheme [24-30], the present quantum eraser relates to the first-order quantum superposition of a single photon in an NMZI [2-4]. Finally, the polarization-basis control in the delayed-choice quantum eraser [32,33] is used to analyze the Bell inequality violation between them (see Sections B~E of Supplementary Materials).

**Results**
*Coherently excited polarization-frequency correlation*

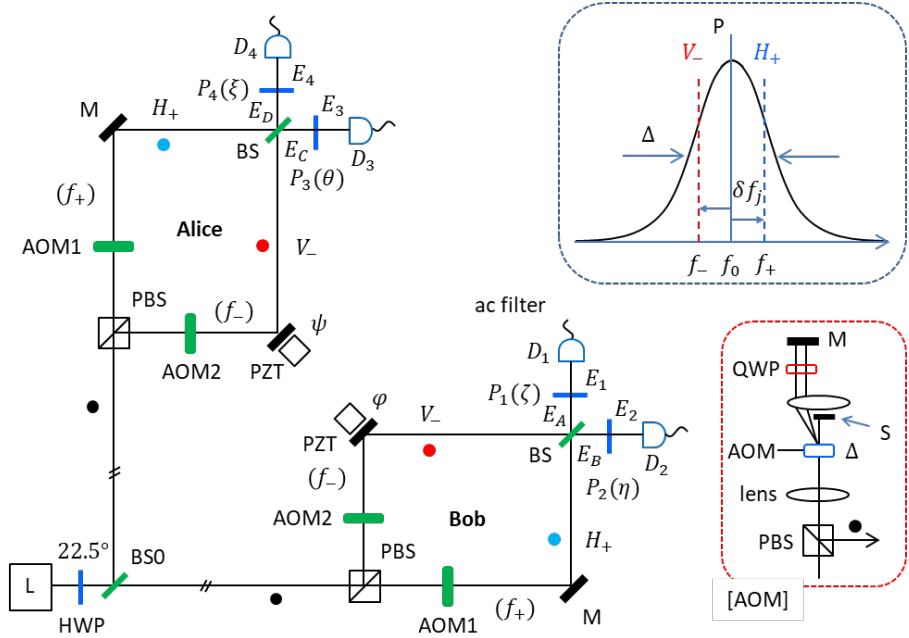

Fig. 1. Schematic of coherently excited nonlocal correlation in a quantum eraser scheme. AOM: acousto-optic modulator, BS0/BS: beam splitter, D: single photon detector, H (V): horizontal (vertical) polarization, HWP: half-wave plate, L: laser, M: mirror, P: polarizer, PBS: polarizing BS, PZT: piezo-electric transducer, QWP: quarter-wave plate. The black, blue, and red dots indicate a single photon. $f_-$ ($f_+$) represents $-\delta f$ ($+\delta f$) detuned frequency from the center frequency $f_0$ by AOMs. Inset (top): AOM generated frequency pair. Inset (bottom): double-pass scheme.

Figure 1 shows a schematic of the coherently excited nonlocal correlation based on a pair of quantum erasers using a continuous wave (cw) laser. In each party, an NMZI is used for the polarization-frequency correlation of a single photon satisfying a quantum eraser scheme [32]. For randomly generated orthogonal polarization bases, a 22.5°-rotated half-wave plate (HWP) is inserted right after the attenuated cw laser. In both NMZIs, a synchronized pair of double-pass AOMs is used for manipulations of the polarization-frequency correlated photons (see Methods): $f_+ - H_+$; $f_- - V_-$. For the mutual coherence between paired photons, the AOMs must be synchronized in each NMZI. This mutual coherence between paired photons is the bedrock of the nonlocal correlation [25,31]. To satisfy the nonlocal condition, the distance between PBS and polarizers in each NMZI is set to be outside the light cone [4, 32]. The single photon condition in each NMZI is post-determined by coincidence detection between NMZIs. Due to Poisson statistics, a ~1 % error by three or more bunched photons is inevitable due to Poisson statistics [32]. In each NMZI, the output photon's polarizations are random by the beam splitter (BS), satisfying basis randomness in local detection. Due to orthogonal polarization bases, however, each NMZI results in no fringe in both output ports [34]. By adding a polarizer in each output port of the NMZI, a quantum eraser scheme is satisfied for the violation of the cause-effect relation, resulting in self-interference fringes [3,4,32]. Finally, this polarizer-controlled measurement scheme in the quantum eraser is adopted to analyze the Bell-inequality violation between NMZIs [22]. Due to the independence between local polarizers, any violation of the Bell inequality witnesses the coherently excited nonlocal correlation for Fig. 1 [20-23]. Due to the fundamental physics of delayed-choice quantum eraser, no difference exists between a single photon [32] and a cw light [33].

*Analysis*

For the coherence approach in Fig. 1 using an attenuated cw laser, each photon's amplitude is represented by $E_0$. Each output photon of the NMZI satisfies quantum superposition between polarization-frequency correlation $|H_+\rangle|f_+\rangle$ and $|V_-\rangle|f_-\rangle$ in a single photon regime, where $f_\pm = f_0 \pm \delta f_j$ (see the bottom Inset). The AOM-generated photon pairs should show an ensemble-based effective-coherence length $l_c = c\Delta^{-1}$ in a frequency scanning regime, where c is the speed of light and $\Delta$ is the AOM's bandwidth. As shown in the bottom Inset, the AOM is set for a double-pass scheme, resulting in $\delta f_j$-independent propagation direction. By the gated coincidence detection of the heterodyne signal between two NMZIs, only doubly-bunched photon cases before the first BS (BS0) are automatically selected for the nonlocal correlation based on orthogonal polarization product bases (different-frequency product bases) (see Methods and Section A of Supplementary Materials) [32].

In each NMZI, the post-selected output photons for the gated heterodyne detection-based coincidence measurements can be represented as:

$$\begin{bmatrix} E_A^j \\ E_B^j \end{bmatrix} = \frac{E_0}{2} e^{i\alpha} e^{i(k_0 r - \omega_0 t)} \begin{bmatrix} H_+ e^{i\delta f_j t_+} - V_- e^{-i\delta f_j t_-} \\ i(H_+ e^{i\delta f_j t_+} + V_- e^{-i\delta f_j t_-}) \end{bmatrix}, \quad (1)$$

$$\begin{bmatrix} E_C^j \\ E_D^j \end{bmatrix} = \frac{E_0}{2} e^{i(k_0 r - \omega_0 t)} \begin{bmatrix} H_+ e^{i\delta f_j t_+} - V_- e^{-i\delta f_j t_-} \\ i(H_+ e^{i\delta f_j t_+} - V_- e^{-i\delta f_j t_-}) \end{bmatrix}. \quad (2)$$

The global phase $e^{i\alpha}$ in Eq. (1) is due to the arbitrary path-length difference from BS0 to both NMZIs. Equations (1) and (2) are for the first-order amplitudes of a single photon, resulting in no fringes due to the PBS-caused distinguishable photon characteristics [34]. Thus, the mean local intensities are $\langle I_A \rangle = \langle I_B \rangle = \langle I_C \rangle = \langle I_D \rangle = I_0$, where $I_0 = E_0 E_0^*$.

By the inserted polarizers in both output ports of the NMZI, Eqs. (1) and (2) are rewritten as:

$$\begin{bmatrix} E_1^j \\ E_2^j \end{bmatrix} = \frac{E_0}{2} e^{i\eta} e^{i(k_0 r - \omega_0 t)} e^{i\delta f_j t_+} \begin{bmatrix} H_+ \cos\zeta - V_- \sin\zeta e^{-i\varphi_j} \\ i(H_+ \cos\eta + V_- \sin\eta e^{-i\varphi_j}) \end{bmatrix}, \quad (3)$$

$$\begin{bmatrix} E_3^j \\ E_4^j \end{bmatrix} = \frac{E_0}{2} e^{i(k_0 r - \omega_0 t)} e^{i\delta f_j t_+} \begin{bmatrix} H_+ \cos\theta - V_- \sin\theta e^{-i\psi_j} \\ i(H_+ \cos\xi + V_- \sin\xi e^{-i\psi_j}) \end{bmatrix}, \quad (4)$$

where $\zeta, \eta, \theta$, and $\xi$ are the rotation angles of the polarizer from the horizontal axis to the counter-clockwise direction [32]. Due to the same polarization projection of the orthogonal polarization bases of a photon onto the corresponding polarizer, the originally non-interacting probability amplitudes of the photon on PBS become now interfered with and result in fringes (see below): a quantum eraser [32]. Regarding the AOM-induced frequency detuning by $\pm\delta f_j$ across the center frequency $f_0$, the phase can be expressed by $\varphi_j = 2\delta f_j \tau_B$ and $\psi_j = 2\delta f_j \tau_A$, where $\tau_k$ $(k = A; B)$ is the time delay of paired photons between NMZI paths. Here, $\tau_{j=A,B}$ can be controlled by either the optical path-length difference of the $f_-$ or the rf pulse (phase) delay for AOM2.

The corresponding intensities of the paired $j^{th}$ photons after the polarizers are represented as follows:

$$I_1^j = \frac{I_0}{4} (H_+ \cos\zeta - V_- \sin\zeta e^{-i\varphi_j})(H_+ \cos\zeta - V_- \sin\zeta e^{i\varphi_j})$$
$$= \frac{I_0}{4} (1 - \sin 2\zeta \cos\varphi_j), \quad (5)$$

$$I_2^j = \frac{I_0}{4} (H_+ \cos\eta + V_- \sin\eta e^{-i\varphi_j})(H_+ \cos\eta + V_- \sin\eta e^{i\varphi_j})$$

$$= \frac{I_0}{4}(1 + sin2\eta cos\varphi_j), \tag{6}$$

$$I_3^j = \frac{I_0}{4}(H_+ cos\theta - V_- sin\theta e^{-i\psi_j})(H_+ cos\theta - V_- sin\theta e^{i\psi_j})$$
$$= \frac{I_0}{4}(1 - sin2\theta cos\psi_j), \tag{7}$$

$$I_4^j = \frac{I_0}{4}(H_+ cos\xi + V_- sin\xi e^{-i\psi_j})(H_+ cos\xi + V_- sin\xi e^{i\psi_j})$$
$$= \frac{I_0}{4}(1 + sin2\xi cos\psi_j). \tag{8}$$

From Eqs. (5)-(8), local measurements of the $j^{th}$ photon show interference fringes, where $cos\varphi_j$ and $cos\psi_j$ relate to $\tau_k$-dependent ensemble decoherence. In each NMZI, such a fringe witnesses the quantum eraser [32], as has also been observed in both SPDC- [9,10] and single-photon cases [4,32]. Here, the mean values of local intensities become insensitive to the NMZI path-length difference due to slow ensemble decoherence in $\langle cos\varphi_j \rangle$ and $\langle cos\psi_j \rangle$.

For the nonlocal correlation, gated heterodyne detection-based coincidence measurements are analyzed for local detectors $D_1$ and $D_4$, satisfying the space-like separation:

$$R_{14}^j = \frac{I_0^2}{16}(H_+ cos\zeta - V_- sin\zeta e^{-i\varphi_j})(H_+ cos\xi + V_- sin\xi e^{-i\psi_j})(cc), \tag{9}$$

where cc is the complex conjugate. Thus, the corresponding mean value $\langle R_{14}(\tau) \rangle (\equiv \frac{1}{N}\sum_{j=1}^{N} R_{14}^j(\tau))$ of Eq. (9) is as follows:

$$\langle R_{14}(\tau) \rangle = \frac{I_0^2}{16} H_+ V_- \langle (-sin\zeta cos\xi e^{-i\varphi_j} + cos\zeta sin\xi e^{-i\psi_j})(c.c.) \rangle,$$
$$= \frac{I_0^2}{16} \langle sin^2(\zeta - \xi) + 2sin\zeta cos\xi cos\zeta sin\xi (1 - cos(\psi_j - \varphi_j)) \rangle, \tag{10}$$

where $\tau = \tau_A - \tau_B$. In Eq. (10), exclusion of the same polarization-basis products by the dc-cut ac-pass filter is the key to the selective measurements, resulting in nonlocal quantum correlation (see Methods and Fig. 2). For the coincidence detection, i.e., $\tau_A = \tau_B$, thus, $\langle R_{14}(0) \rangle = I_0^2 sin^2(\zeta - \xi)/16$ is obtained for the joint-parameter relation of local polarizers: This expression is the quintessence of the nonlocal quantum feature, showing the inseparable product basis between $\zeta$ and $\xi$ [23]. As the time delay $\tau$ increases, the deterioration of the nonlocal quantum feature increases. For $\tau \gg \Delta^{-1}$, Eq. (10) turns out to be classical (see Fig. 3): $\langle R_{14}(\tau \gg \Delta^{-1}) \rangle = I_0^2(sin^2\zeta cos^2\xi + cos^2\zeta sin^2\xi)/16$. Likewise, $\langle R_{23}(\tau) \rangle = \langle R_{14}(\tau) \rangle$ is also satisfied (not shown).

Similarly, the following correlation relation is obtained between detectors $D_1$ and $D_3$ (see Section B of the Supplementary Material):

$$\langle R_{13}(0) \rangle = \langle R_{24}(0) \rangle = \frac{I_0^2}{4} sin^2(\zeta + \theta), \tag{11}$$

where Eq. (11) cancels Eq. (10) in the joint parameter relation (see Fig. 2). Thus, the sum of $\langle R_{14}(0) \rangle$ and $\langle R_{13}(0) \rangle$ ($\langle R_{23}(0) \rangle$ and $\langle R_{24}(0) \rangle$) shows the classical feature without the joint parameter relation, which is equal to $R_{14}(\tau \gg \Delta^{-1})$ (see Fig. 3 and Section B of the Supplementary Material and Fig. 3). Thus, the coherently excited nonlocal quantum correlation is successfully derived for Fig. 1 via polarization-frequency correlation control and gated heterodyne detection-based coincidence measurements.

Figure 2 shows numerical calculations of the analytical solutions in Eqs. (5)-(11). The upper panels are for the quantum features of Bell inequality violation at $\tau = 0$, i.e., $\psi_j = \varphi_j$. As shown in the upper right panel, coincidence detection-based two-photon correlation coefficients are calculated from the four points. As a result, $E(\alpha, \beta) = [R_{14}(\alpha, \beta) + R_{23}(\alpha, \beta) - R_{13}(\alpha, \beta) - R_{24}(\alpha, \beta)] = 0$ is obtained from the lower left point of the blue ($R_{14}$) and red ($R_{14}$) dotted curves, where $\alpha = \zeta, \eta$ and $\beta = \xi, \theta$ [21,23]. Using those calculated E values for $(\alpha, \alpha', \beta, \beta') = (0°, 45°, 22.5°, 67.5°)$, the Bell parameter S is calculated to be violated the local hidden variable theory bounded by S=2 (see Section C of the Supplementary Materials): $S(\alpha, \alpha', \beta, \beta') = |E(\alpha, \beta) + E(\alpha', \beta) - E(\alpha, \beta') + E(\alpha', \beta')| = 2\sqrt{2}$. Thus, the Bell inequality violation is numerically confirmed for derived coherence solutions of the coherently excited nonlocal correlation in Fig. 1.

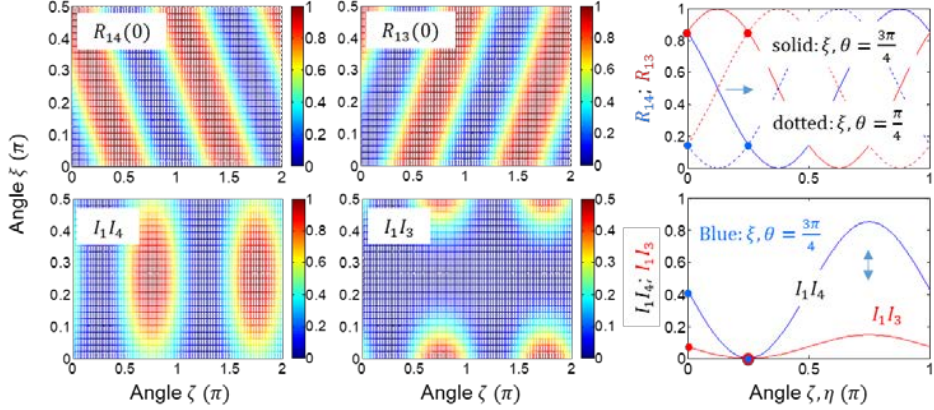

Fig. 2. Numerical calculations for Fig. 1. (Upper panels) Nonlocal quantum feature of Eqs. (10) and (11) for $\tau = 0$. (Lower panels) Intensity product of $I_1I_4$; $I_1I_3$. (Right-end column) Blue: $R_{14}$ $(= R_{23})$; $I_1I_4$ $(= I_2I_3)$. Red: $R_{13}$ $(= R_{24})$; $I_1I_3$ $(= I_2I_4)$. For the bottom panels, no dc-cut is applied [32].

On the contrary to the upper panels of the quantum features obtained by the gated heterodyne detection-based coincidence measurements, the lower panels of Fig. 2 are for the corresponding classical features without the gated detection. For this, both locally measured intensity products, $I_1I_4$ $(= I_2I_3)$ and $I_1I_3$ $(= I_2I_4)$ are calculated for E values, where $I_1I_4$ and $I_1I_3$ are represented by blue and red curves in the lower right panel, respectively. Each curve is for $(\xi, \theta) = (22.5°, 67.5°)$. Unlike the upper right panel, the Bell parameter S of the lower right panel does not violate the classical upper bound (see Section D of the Supplementary Materials): S=0. Thus, the present coherence approach is successful to excite the nonlocal quantum feature of Bell inequality violation, where the joint-parameter relations in Eqs. (10) and (11) are the origin of the so-called quantum steering [35]. However, it is quite interesting that Eqs. (10) and (11) are seemingly contradictory to the violation of local realism simply because each photon pair is coherent between NMZIs to preserve their predetermined realism no matter how far they are apart even for $\tau > \Delta^{-1}$. This seeming contraction is of course resolved with the conceptual definition of information meaningful in an ensemble, as discussed in superluminal debates [36]. In other words, the effective coherence of the photons' spectral bandwidth should be the guideline of the local realism, even though individual coherence has to be persistent for the nonlocal correlation.

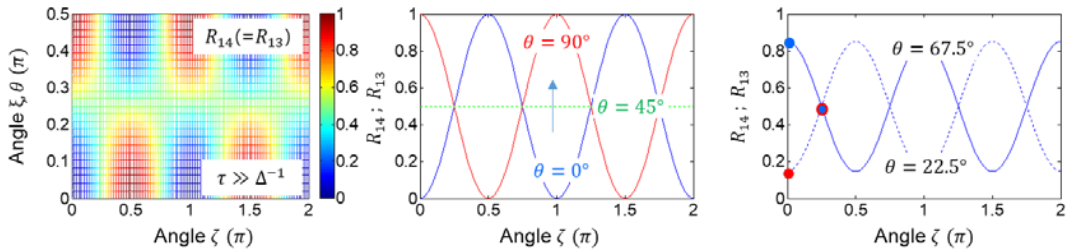

Fig. 3. Numerical calculations for Eq. (10) and (11) at $\tau \gg \Delta^{-1}$. $\xi = \theta$ is satisfied independently for $\zeta$.

Figure 3 shows the deterioration of the quantum feature as the time delay $\tau$ increases. At $\tau \gg \Delta^{-1}$, the nonlocal quantum feature completely disappears, resulting in the classical feature without satisfying the joint parameter relation. Compared with the local intensity product $I_1I_4$ $(\neq I_1I_3)$ in Fig. 2 (bottom panels), Fig. 3 shows a bit different feature, where $R_{13} = R_{14}$ at $\tau \gg \Delta^{-1}$. The Bell parameter S does not violate the Bell inequality condition, either (see Section E of the Supplementary Materials): S=0. The loss of quantum correlation in Fig. 3 is due to the ensemble decoherence of all measured events for $\tau \neq 0$. In other words, mutual coherence between individually paired photons is the bedrock of the nonlocal quantum feature.

**Discussion**
Do those nonlocal quantum features of Eqs. (10) and (11) violate local realism, i.e., the hidden variable theory [18-21]? The heart of local realism is the pre-determinism by classical laws such as in coherence optics. Thus, local realism states that any action in one party cannot affect the physical entity in the other party apart from a space-like separation. The quantum feature derived in Eq. (10) requires mutual coherence between paired photons for measurement-modified product bases, as presented in Fig. 2 (see the upper panels). Within this coherence, the joint measurements confirm that the local action of one party affects measurement results in the other party, as shown in the upper right panel of Fig. 2. Not until measured by coincidence detection [23] or informed of basis choice [37,38], however, there is no way to know one's state in the other party. As the retrieval of the predetermined coherence feature is the essence of the quantum eraser [32], the gated detection of the paired photons is for a check-up process of the preexisting correlation in a selective measurement way. In other words, the quantum loophole of the locality might be meaningful for an ensemble due to the retrieval of the preexisting mutual coherence for the individual basis, as shown in Figs. 2 and 3. However, this coherence retrieval is not for collective but individual in photon pair measurements, as discussed in Franson correlation [31]. Thus, the nonlocal realism should be effective to an ensemble, otherwise shows a quantum illusion for the predetermined correlation. Without mutual coherence between individually paired photons, nothing happens. To avoid of this contradiction, the nonlocal realism must be meaningful with respect to the information transfer based on an ensemble decoherence, which also defines the space-like separation.

Regarding the selective measurement by the gated heterodyne detection, the 50 % event loss is to accomplish quantum superposition between the orthogonal polarization-product bases. This reduced product bases between paired photons results in increased correlation beyond the classical limit. In the Franson correlation, the path-basis products (S-S and L-L) are manipulated for the same 50 % event loss [31]; In the quantum eraser, the 50 % event loss is also necessary for the reversal of quantum property [32, 39]. These selective measurements relate to the same reduced tensor products between bipartite particles. Thus, the nonlocal realism is illusive due to the modification of measurement events. The selective measurement-based nonlocal quantum features discussed in Eqs. (10) and (11) are closely related to the entanglement purification or distillation process applied to decoherent or mixed states [40,41]. In a macroscopic regime, it is also possible to reach the same quantum feature as in the single-photon case (discussed elsewhere), because the quantum eraser works even for a cw regime [33].

**Conclusion**
Using coherent manipulations of polarization-frequency correlated photons in an NMZI for a quantum eraser, the nonlocal correlation between space-like separated NMZI output photons was analyzed and numerically demonstrated for the Bell-inequality violation. The manipulation of polarization-frequency correlation was conducted by a pair of synchronized AOMs in an NMZI. The role of gated detection of heterodyne signals is to selectively choose orthogonal polarization-product bases via a dc-cut ac-pass filter, resulting in the joint-parameter relation between quantum erasers. Unlike the conventional understanding of the nonlocal correlation limited to quantum particles such as entangled photon pairs from SPDC, the present coherence approach using an attenuated laser was contradictory to the common understanding that nonlocal correlation cannot be achieved by any classical means. With gated detection-caused measurement-basis modifications for heterodyne signals, however, the proposed coherence scheme was successful for nonlocal quantum features in a space-like separation regime, satisfying the Bell inequality violation. Thus, the nonlocal quantum feature was rooted in mutual coherence between paired photons via selective polarization-product bases in a gated detection regime. As a result, the violation of local realism must be effective for an ensemble in terms of information transfer to avoid the same debate in the superluminal light propagation. For the nonlocal correlation, however, mutual coherence between paired photons in each measurement event was a prerequisite to keep the predetermined photon correlation, even beyond the ensemble decoherence-defined space-like separation.

**Methods**

For the gated detection in Fig. 1, the mean photon number is set to be extremely low, satisfying the incoherence condition between consecutive single photons. For this, a mean consecutive photon distance is set to be much longer than the laser's coherence length [42]. For the coincidence measurements, however, only Poisson-distributed doubly-bunched photons are post-selected for the test of the nonlocal quantum feature, whose generation rate is ~ 1 % with respect to single photons [42]. Unlike SPDC-based entangled photons, whose frequencies are oppositely detuned across the center frequency $f_0$, coherent photons from L have no such detuning relation due to the cavity optics. The role of paired AOMs is to generate $\pm \delta f_j$ (frequency)-correlated photon pairs like the SPDC case (see the top Inset of Fig .1). For this, the laser linewidth must be much narrower than the bandwidth of AOMs. For the random detuning, the AOMs are synchronously scanned for its bandwidth, whose scan speed is adjusted to satisfy the randomly detuned photon pairs.

In Fig. 1, a doubly bunched photon pair can be either split into both NMZIs or bunched together into each NMZI by BS0 at an equal chance. The bunched photon case is automatically dropped out by the definition of coincidence measurements of the heterodyne signals between two NMZIs. For the split photon case, each NMZI output photon can be either vertically or horizontally polarized at an equal chance, where the blue and red dots in Fig. 1 represent the same probability amplitude of the single photon (black dot). The bunched photon case into either NMZI output port is not allowed for this single-photon input or the heterodyne detection. The goal of the gated detection of the heterodyne signals caused by AOM-induced $\pm \delta f_j$ is to selectively choose an orthogonally polarized photon pair in the delay-time τ domain to separate other pairs in the free running time 't.' Here, it should be noted that no beating signal is induced for the same polarization-based photon pairs, which must be avoided for selective measurements. For this, a dc-cut ac-pass filter should be inserted for the coincidence measurements. Thus, the success rate of the gated detection-based coincidence measurements is one out of four in the probability amplitude (see Section A of Supplementary Materials). For the NMZI output photons, only 50 % of measurement events are chosen by the ac-pass filter. This selective measurement is the quintessence of the nonlocal correlation using coherent photons from an attenuated laser.

**Acknowledgments**

This work was supported by the ICT R&D program of MSIT/IITP (2023-2021-0-01810), the development of elemental technologies for an ultrasecure quantum internet. BSH also acknowledges that this work was supported by the GIST Research Project in 2023.

**Reference**
1. Bohr, N. in Quantum Theory and Measurement, Wheeler, J.A. & Zurek, W.H. Eds. (Princeton Univ. Press, Princeton, NJ), pages 949, 1984.
2. Wheeler, J. A. in Quantum Theory and Measurement, J. A. Wheeler and W. H. Zurek eds (Princeton University Press, 1984), pp. 182-213.
3. Scully, M. O., Englert, B.-G. & Walther, H. Quantum optical tests of complementarity, *Nature* **351**, 111-116 (1991).
4. Jacques, V., Wu, E, Grosshans, F., Treussart, F., Grangier, P., Aspect, A. & Roch, J.-F. Experimental realization of Wheeler's delayed-choice Gedanken Experiment. *Science* **315**, 966-978 (2007).
5. Manning, A. G., Khakimov, R. I., Dall, R. G. & Truscott, A. G. Wheeler's delayed-choice gedanken experiment with a single atom. *Nature Phys.* **11**, 539-542 (2015).
6. Aharonov, Y. & Zubairy, M. S. Time and the Quantum: Erasing the Past and Impacting the Future. *Science* **307**, 875-879 (2005).
7. Tang, J.-S., Li, Y.-L., Xu, X.-Y., Xiang, G.-Y., Li, C.-F. & Guo, G.-C. Realization of quantum Wheeler's delayed-choice experiment. *Nature Photon.* **6**, 600-604 (2012).
8. Ionicioiu, R. & Terno, D. R. Proposal for a Quantum Delayed-Choice Experiment. *Phys. Rev. Lett.* **107**, 230406 (2011).

9. Ma, X.-S., Kofler, J. & Zeilinger, A. Delayed-choice gedanken experiments and their realizations. *Rev. Mod. Phys.* **88**, 015005 (2016).
10. Herzog, T. J., Kwiat, P. G., Weinfurter, H. & Zeilinger, A. Complementarity and the quantum eraser. *Phys. Rev. Lett.* **75**, 3034-3037 (1995).
11. Ionicioiu, R., Jennewein, T., Mann, R. B. & Terno, D. R. Is wave-particle objectivity compatible with determinism and locality? *Nature Communi.* **5**, 4997 (2014).
12. Wang, K., Xu, Q., Zhu, S. & Ma, X.-S. Quantum wave-particle superposition in a delayed-choice experiment. *Nature Photon.* **13**, 872-877 (2019).
13. Christensen, B. G. *et al.*, Detection-loophole-free test of quantum nonlocality, and applications. *Phys. Rev. Lett.* **111**, 130406 (2013).
14. Aspect, A. Closing the door on Einstein and Bohr's quantum debate. *Physics* **8**, 123 (2015).
15. Giustina, M. *et al.*, Significant-loophole-free test of Bell's theorem with entangled photons. *Phys. Rev. Lett.* **115**, 250401 (2015).
16. Hensen, B. *et al.*, Loophole-free Bell inequality violation using electron spins separted by 1.3 kilometres. *Nature* **526**, 682-686 (2015).
17. Abellan, C. *et al.*, Challenging local realism with human choices. *Nature* **557**, 212-216 (2018).
18. Einstein, A., Podolsky, B. & Rosen, N. Can quantum-mechanical description of physical reality be considered complete? *Phys. Rev.* **47**, 777–780 (1935).
19. Wiseman, H., Death by experiment for local realism. *Nature* **526**, 649-650 (2015).
20. Bell, J. S. On the Einstein Podolsky Rosen paradox. *Physics* **1**, 195–200 (1964).
21. Clauser, J. F., Horne, M. A., Shimony, A. & Holt, R. A. Proposed experiment to test local hidden-variable theories. *Phys. Rev. Lett.* **23**, 880-884 (1969).
22. Aspect, A., Grangier, P. & Roger, G. Experimental realization of Einstein-Podolsky-Rosen-Bohm Gedankenexpriment: A new violation of Bell's inequalities. *Phys. Rev. Lett.* **49**, 91-94 (1982).
23. Weihs, G., Jennewein, T., Simon, C., Weinfurter, H. & Zeilinger, A. Violation of Bell's inequality under strict Einstein locality. *Phys. Rev. Lett.* **81**, 5039-5042 (1998).
24. Franson, J. D. Bell inequality for position and time. *Phys. Rev. Lett.* **62**, 2205-2208 (1989).
25. Kwiat, P. G., Steinberg, A. M. & Chiao, R. Y. High-visibility interference in a Bell-inequality experiment for energy and time. *Phys. Rev.* A **47**, R2472–R2475 (1993).
26. Lima, G., Vallone, G., Chiuri, A., Cabello, A. & Mataloni, P. Experimental Bell-inequality violation without the postselection loophole. *Phys. Rev. A* **81**, 040101(R) (2010).
27. Carvacho, G. *et al.*, Postselection-loophole-free Bell test over an installed optical fiber network. *Phys. Rev. Lett.* **115**, 030503 (2015).
28. Marcikic, I., de Riedmatten, H., Tittle, W., Scarani, V., Zbinden, H. & Gisin, N. Time-bin entangled qubits for quantum communication created by femtosecond pulses. *Phys. Rev.* A **66**, 062308 (2002).
29. Aerts, S., Kwiat, P., Larsson, J.-Å. & Żukowski, M. Two-photon Franson-type experiments and local realism. *Phys. Rev. Lett.* **83**, 2872-2875 (1999).
30. Cuevas, A. *et al.*, Long-distance distribution of genuine energy-time entanglement. *Nature Communi.* **4**, 2871 (2013).
31. Ham, B. S. The origin of Franson-type nonlocal correlation. arXiv:2112.10148v3 (2023).
32. Kim, S. & Ham, B. S. Observations of the delayed-choice quantum eraser using coherent photons. Sci. Rep. **13**, 9758 (2023).
33. Ham, B. S. Observations of the delayed-choice quantum eraser in a macroscopic system. arXiv:2205.14353v3 (2022).
34. Hardy, L. Source of photons with correlated polarizations and correlated directions. Phys. Lett. A **161**, 326-328 (1992).
35. Uola, R., Costa, A. C. S., Nguyen, H. C. & Gühne, O. Quantum steering. Rev. Mod. Phys. **92**, 015001 (2020).
36. Wang, L. J., Kuzmich, A. & Dogariu, A. Cain-assisted superluminal light propagation. Nature **406**, 277-279 (2000).

37. Bouwmeester, D., Pan, J.-W., Mattle, K., Eibl, M., Weinfurter, H. & Zeinlinger A. Experimental quantum teleportation. Nature **390**, 575-579 (1997).
38. Marcikic, I., de Riedmatten, H., Tittel, W., Scarani, V., Zbinden, H. & Gisin, N. Time-bin entangled qubits for quantum communication created by femtosecond pulses. Phys. Rev. A **66**, 062308 (2002).
39. Ham, B. S. A coherence interpretation of nonlocal realism in the delayed-choice quantum eraser. arXiv: 2302.13474v3 (2023).
40. Bennett, C. H., Brassard, G., Popescu, S., Schumacher, B., Smolin, J. A. & Wooters, W. K. Purification of noisy entanglement and faithful teleportation via noisy channels. Phys. Rev. Lett. **76**, 722-725 (1996).
41. Pan, J.-W., Simon, C., Brukner, C. & Zeilinger, A. Entanglement purification for quantum communication. Nature **410**, 1067-1070 (2001).
42. Kim, S. and Ham, B. S. Revisiting self-interference in Young's double-slit experiments. Sci. Rep. **13**, 977 (2023).